\documentstyle[twocolumn,prl,aps,psfig]{revtex}
\begin{document}
\draft
\title{Collective Excitations, NMR, and Phase Transitions in Skyrme
Crystals}
\author{R. C\^{o}t\'{e}$^{1}$, A.H. MacDonald$^{2}$, Luis Brey$^{3}$,
H.A. Fertig$^{4}$, S.M. Girvin$^{2}$ and H.T.C. Stoof$^{5}$}
\address{$^{1}$ D\'{e}partement de Physique
Universit\'{e} de Sherbrooke, Sherbrooke,
Qu\'{e}bec, Canada J1K-2R1}
\address{$^{2}$ Department of Physics, Indiana University,
Bloomington IN 47405}
\address{$^{3}$ Instituto de Ciencia de Materiales (CSIC),
Universidad Aut\'{o}noma C-12, 28049 Madrid, Spain}
\address{$^{4}$ Department of Physics and Astronomy, University of
Kentucky, Lexington KY 40506-0055}
\address{$^{5}$ Institute for Theoretical Physics, University of
Utrecht, P.O. Box 80.006, 3508 TA Utrecht, The Netherlands}
\date{\today}
\preprint{IUCM97-008;CRPS-97-11}
\maketitle

\begin{abstract}

At Landau level filling factors near $\nu =1$, quantum Hall ferromagnets 
form a Skyrme crystal state with quasi-long-range 
translational and non-collinear magnetic order.  
We develop an effective low energy theory which 
explains the presence in these systems of magnetic excitations at low energies 
below the Larmor gap ($\Delta$) and which predicts a dramatic enhancement 
of the nuclear spin relaxation rate by a factor of $10^3$.  
The effective theory predicts
a rich set of quantum and classical phase transitions.  
Based in part on accurate time-dependent Hartree-Fock calculations
of the ordered state collective excitation spectrum, we 
discuss aspects of the $T-\nu-\Delta$ Skyrme crystal phase diagram.

\pacs{74.60Ec;74.75.+t}

\end{abstract}

At Landau level filling factor $\nu =1/m$ (with $m$ an odd integer), 
the ground state of a two-dimensional electron system (2DES) is an 
incompressible strong ferromagnet\cite{dassarmabook}, {\it i.e.}, 
it has complete spin alignment even in the limit of vanishing Zeeman 
coupling strength and it has a gap for charged excitations. 
Quantum Hall ferromagnets have the unusual 
property\cite{sondhi,leekane,moon} that their topologically non-trivial 
spin texture excitations (skyrmions) carry charge.
This property profoundly affects their physics.  In
particular, the {\it ground} state at filling factors slightly away
from  $\nu =1/m$ contains a finite density of skyrmions and these are 
expected\cite{skyrmlat,hf,tsvelik} to crystallize, at least when they are 
sufficiently dilute.  

In an external field the spin-wave modes of a collinear ferromagnet 
have an excitation gap equal to the Zeeman
energy.  On the other hand, a non-collinear magnet can have two  
Goldstone modes\cite{sachdev} and one of these can remain
gapless in an external field.  The Skyrme crystal state has 
non-collinear magnetic order\cite{skyrmlat}.
A single skyrmion spin texture has its spins aligned with the 
Zeeman field at infinity, reversed 
at the center of the skyrmion, and has
non-zero XY spin components at intermediate distances which have 
a vortex like configuration.\cite{sondhi,hf}
The classical (or quantum mean-field)
energy of a skyrmion is independent of the 
angle $\varphi$ which defines the global orientation of the XY spin
components.  This extra $U(1)$ degree of freedom for a single Skyrmion 
leads to broken symmetry in the crystal ground state and
hence to a spin wave mode which remains gapless in the presence
of the Zeeman field.  The existence of this gapless
spin mode (presumably in overdamped form in a Skyrme {\it liquid} state)
dramatically alters the low temperature physics, manifesting itself
in both rapid nuclear spin relaxation\cite{barrett} and giant
apparent specific heat\cite{bayot}.

In this Letter we address the impact of thermal and quantum
fluctuations on the physics of Skyrme crystals.  We propose a rich zero
temperature phase diagram in which quantum fluctuations destroy the magnetic
order of the Skyrme crystal for some values of $\nu$ and Zeeman coupling
strength.  Where order exists in the ground state, we estimate the finite
temperature Kosterlitz-Thouless phase transition temperatures. Our analysis
is based on a boson Hubbard model, which we argue describes the low-energy
physics of this system, and on time-dependent Hartree-Fock approximation
(TDHFA) calculations of the collective modes of the Skyrme crystal which 
accurately capture the intricate microscopic physics of the crystal state 
and fix the Hubbard model parameters.

The simplest symmetry-allowed orientation-dependent coupling between
neighboring skyrmions is of the form 
$V= J\sum_{\langle ij\rangle}\cos\left(\varphi_i -\varphi_j\right)$.
For typical Zeeman coupling strengths and $\nu$ not too close to
$1$, detailed Hartree-Fock calculations\cite{skyrmlat} show that the Skyrme
crystal is a square lattice with opposing 
orientations ( $\varphi_i = \varphi_j + \pi$)
for skyrmions on opposite sublattices. 
The low-energy effective Hamiltonian we propose 
is motivated by microscopic
theory from which it follows\cite{hcmskyrm} that, in addition to the
positional degrees of freedom, skyrmions
carry an integer-valued internal
quantum number ($K$) which is the number of flipped spins (relative to 
the maximally polarized state of the same charge). 
Quantum states of definite $K$ are related to the classical states of
definite orientation by\cite{hcmskyrm,nayak}
$|K\rangle \propto \int d\varphi \exp(iK\varphi)|\varphi\rangle$, so that
for each skyrmion $K_i$
and $\varphi_i$ are canonically conjugate.
Thus our effective model is equivalent to  
a boson Hubbard model\cite{fisher} in which 
the boson number on the $i$th site is mapped to the number of 
flipped spins in the $i$th skyrmion.  
In this language the orientation-dependent interaction term 
corresponds to boson hopping and favors long range boson coherence. 

We limit our attention for the moment to states where the skyrmions have
crystallized.  At low energies, phonon excitations of the
lattice decouple\cite{tobepub} from the spins and can be treated separately.
Because of the Larmor gap and the 
incompressibility gap for particle-hole excitations,
fluctuations in $K$ are the only relevant low-energy 
degrees of freedom, if the skyrmion positions are fixed.
The energy of an isolated skyrmion in state $K_i$ can be written
in the form\cite{hf}
$\epsilon_{K_i}=U_{K_i}+g^{*}\mu _{B}BK_i$,
where $g^{*}$ is the host semiconductor g-factor.
Non-linearity in the function $U(K)$ maps to on-site boson interactions.
In the Zeeman coupling term on the right hand side of this equation,
$- \tilde g \equiv -g^{*} \mu _{B}B$,
plays the role of a chemical potential for the bosons.   
(In typical experimental systems $\tilde g \sim 0.015 (e^2/\epsilon \ell)$.)
Making a quadratic approximation for $U_K$ we are led to an effective
Hamiltonian describing boson fluctuations for states with $K$ near  
$\langle {\hat K} \rangle   \equiv K^0 (\tilde g,\nu) $:
\begin{equation}
H = U\sum_i(\hat K_i - K^0)^2 +
 J\sum_{\langle ij\rangle}\cos\left(\varphi_i -\varphi_j\right).
\label{skeng}
\end{equation}
For $J\gg U$, boson hopping dominates over interactions  
so that the bosons condense into a supersolid phase with long
range phase coherence.
A linear spin wave approximation for this model yields
a gapless Goldstone mode with dispersion
\begin{equation}
E({\bf k})=\left\{ 4UJ\left[2-\cos (k_{x}a)-\cos (k_{y}a)\right]\right\}^{1/2}
\label{mode}
\end{equation}
However, unless $2 K^{0}$ is an odd integer, 
quantum fluctuations will destroy the classical ordered state
and produce a gapped `insulating' state\cite{fisher}
when $U \gg J$. 

The parameters of this model can be fixed using microscopic
Hartree-Fock calculations\cite{skyrmlat}.  From the ground state
calculation we can evaluate the number of reversed
spins per charge, $\langle {\hat K} \rangle $.
In the harmonic approximation, the boson interaction parameter $U$ 
is given by the inverse boson compressibility:
\begin{equation}
U = - \frac{1}{2} 
\left(\frac{\partial \langle K(\tilde g,\nu)\rangle }
{\partial \tilde g } \right)^{-1}.
\label{ueqn}
\end{equation}
The boson hopping parameter $J$ can then be
extracted from the collective mode spectrum
which we compute by locating poles of the
time-dependent Hartree-Fock approximation (TDHFA) 
spin and density response functions, 
$
\chi_{\mu\nu}\left( {\bf k}+{\bf G},{\bf k}+{\bf G}^{\prime }
,\omega \right).
$
The TDHFA calculation is facilitated by simplifications which arise
from the restricted Hilbert space of the lowest
Landau level\cite{cote1,tobepub}. 
The collective mode spectrum in the extended BZ
of the square lattice Skyrme crystal
is illustrated in Fig.~\ref{collective}.
The two gapless modes are the phonon mode and the new
spin-wave collective mode discussed above.
The dashed lines show the same modes folded into
the smaller BZ of the magnetic lattice. 
As $k \rightarrow 0$, the higher energy of the two-zone
folded modes evolves into the
Larmor mode (global precession of electronic spins around the
Zeeman axis) which must occur at $\hbar\omega_0 = g^{*} \mu_B B$.
The fact that the (new) spin wave mode at $\bf k $ equal to a
reciprocal lattice vector of the magnetic lattice is
equivalent to the $\bf k=0$ Larmor mode can be understood in terms
of the opposite orientations of skyrmions on the two
sublattices.\cite{tobepub}
Because of the inhomogeneous magnetic order of the Skyrme crystal
state, spin and density response functions couple at general points 
in the BZ and poles occur in all response functions at each 
collective mode. 
Fig.~\ref{collective} confirms that at low energies
the only relevant degrees of freedom are skyrmion orientations
and locations, and that these 
give rise to well separated spin-wave and phonon modes.
As $k\rightarrow 0,$ the phonon mode is lowest in energy 
and has the characteristic $k^{3/2}$ magnetophonon dispersion.
We estimate $J$ by requiring Eq.~(\ref{mode}) to reproduce the 
spin-wave velocity of the microscopic TDHFA calculations.
In Fig.~\ref{JUKfig} we plot the boson Hubbard model
parameters ($K^0$, $U$ and $J$) for square lattice Skyrme crystals at a
series of $\tilde g$ values and for
filling factors $\nu =1.05$, $\nu = 1.1$, $\nu = 1.15$, and $\nu =1.2$.

Even the $T=0$, $\nu - \tilde g$ phase diagram of the Skyrme crystal
states is very rich.  Fig.~\ref{phasediagram} summarizes
conclusions (some still qualitative) we have drawn from 
present calculations.
The line of primary importance in this phase diagram is the 
dark solid line which delimits\cite{latcaveat} the stability
region of the square lattice.
In our calculations the instability of the
square lattice is indicated by vanishing long-wavelength
magnetophonon energies.  The square lattice shear constant 
is negative for classical electrons and this structure is 
therefore expected\cite{tobepub} to become unstable 
when skyrmion orientation-dependent interactions become weak.
The square lattice is stable in the 
small $\tilde g$, large $|\nu -1|$ region where 
the skyrmion size is comparable to the crystal lattice
constant.  At low skyrmion density, the critical value of
$\tilde g$ is expected to vary as $|\nu-1|^{3/2}$
corresponding\cite{tobepub} to a fixed ratio of skyrmion size to 
Skyrme crystal lattice constant.
We have not yet completed a systematic survey of possible crystal 
structures outside this region where, in any event,
skyrmion-disorder interactions will frequently 
play a larger role than orientation-dependent skyrmion-skyrmion 
interactions.  Inside the square Skyrme crystal
portion of the diagram our estimates place
$J/U > 1$ except at large $\nu -1$.  We therefore expect to have a ground
state with translational and magnetic order in the shaded portion 
of the phase diagram which excludes regions where $K^0$ is near an integer,
$J/U < 1$, and we expect magnetic-order to be destroyed
by quantum fluctuations.  This phase diagram is subject to
quantitative alteration when quantum fluctuations
in the skyrmion positions become large at large $|\nu -1|$. 
Outside of the square-lattice portion of the phase diagram,
Skyrmion orientations will be more weakly coupled and we expect 
quasi-long-range magnetic order to be rare. 

\null We can extract a rough upper bound
for the Kosterlitz-Thouless critical temperature associated
with the loss of quasi-long-range magnetic order:
$T_{KT} = (\pi/2 k_B) \rho_s \sim 0.008 (e^2/\epsilon\ell) \sim 1{\rm K}$.
Here $\rho_s = J$ is the spin stiffness for the classical
XY model obtained when $U=0$ and the numerical estimate is for
typical magnetic fields $\sim 10 \ {\rm Tesla}$.  
This temperature
should be compared with the classical melting temperature of the crystal
which can be estimated using KTHNY\cite{kthny} theory.  For 
elastically isotropic systems with long  
range interactions, $T_{M} = (b^2 \mu / 4 \pi)$ where
$\mu$ is the shear constant and $b$ the lattice constant of the crystal.
Neglecting the elastic anisotropy of the square
lattice\cite{lanlif,tobepub}, we estimate $\mu$ for
the Skyrme crystal from the long-wavelength dispersion of
the magnetophonon mode
$
\hbar \omega (k) \approx \left(2\pi\mu \ell e^2/\epsilon\right)^{1/2}
(k \ell)^{3/2}
\null .$
These considerations lead to melting temperatures
at $\tilde g =0.015 (e^2/\epsilon \ell)$
of $5.23,3.63$ and $2.51$ (in units of
$10^{-3}(e^2/\epsilon\ell) \sim 0.1$K) for $\nu = 1.10, 1.15$ and $1.20$ 
respectively.  Therefore we expect that
the melting transition will occur at a lower temperature than 
the magnetic transition.  Dislocations in the 
`tetratic' fluid state frustrate the skyrmion orientational order, 
so the spins presumably disorder at the same temperature in a
single transition.\cite{tobepub}  
For classical electrons the melting temperature 
of the electron crystal is lower than the
KTHNY estimate by a factor of $\sim 2$; 
quantum positional fluctuations\cite{dsfisher} cause a further reduction.
Since orientational order is responsible for the positive $T=0$ 
shear modulus of the square Skyrme crystal,
thermal and quantum fluctuations of the XY order are likely to
lead to an even larger reduction in the present case.

At $\nu =1$, spin-relaxation in a quantum Hall ferromagnet
is activated because of the Larmor mode gap.  For $|\nu-1|
\ne 0$, however, both phonon and spin-wave Goldstone modes
of the Skyrme crystal can relax nuclear spins.
The (spatially averaged) relaxation rate has a Korringa temperature
dependence, 
\begin{equation}
\frac{1}{T_1} \propto  \frac{k_{\rm B}
T}{\hbar\Omega} \sum_{\bf G}\int d^2k\, 
\chi_{+-}''({\bf k+G},{\bf k+G},\Omega)
\label{eq:t1}
\end{equation}
where $\Omega$ is the nuclear resonance frequency.
$\chi_{+-}''$ has contributions proportional\cite{tobepub} to
$\delta(\hbar \Omega - \epsilon_j(\vec k))$
where $\epsilon_j(\vec k)$ is one of the collective mode dispersions
shown in Fig. 1.  At long wavelengths neither phonon nor spin-wave 
excitations change the total electronic spin $S_z$ so that there is no
contribution to $T_1^{-1}$ from the ${\bf G}=0$ term in Eq.~\ref{eq:t1}.
However both change $S_z$ locally,
and contribute ${\bf G} \ne 0$ terms to Eq.~\ref{eq:t1} which  
can be extracted from our TDHFA calculations.
We find that\cite{tobepub} 
\begin{eqnarray}
t_{sw}^{-1} &=& \frac{ |\nu -1| X_{sw} (\hbar \omega_c)^2 }{4 \pi U J } \cr
t_{ph}^{-1} &=& \frac{2 X_{ph} (\hbar \omega_c)^2}
{ 3 (4 \pi |\nu -1| \hbar \Omega k_B T_M (e^2/\epsilon \ell))^{2/3}} 
\label{eq:t1result} 
\end{eqnarray} 
where $t_{sw}^{-1}$ and $t_{ph}^{-1}$ are the spin-wave and 
phonon contributions to the relaxation rate
normalized to the Korringa expression for the same
2DES at $B=0$, and $X_{sw}$ and $X_{ph}$ are 
numerical constants.  (For $\nu =1.10$ and $\tilde g = 0.010 (e^2/\ell)$ 
our detailed numerical calculations yield $X_{sw} = 13$
and $X_{ph} = 1.9$.)  We expect that, in practice, disorder
pinning of the Skyrmion lattice 
will gap the phonon spectrum and suppress $t_{ph}^{-1}$.
The linear relationship we find 
between $t_{sw}^{-1}$ and $|\nu -1|$ is in agreement with experiment.
Using parameter values from Fig. 2,
Eq.~\ref{eq:t1result} implies that for $|\nu -1| = 0.1 $, the 
spin-relaxation rate should typically be $\sim 10^3$ times faster 
than for a 2D Fermi gas system at $B=0$, in rough quantitative 
agreement with the experimental results of
Barrett {\it et al.}\cite{barrett,tobepub}.
This enormous enhancement in $1/T_1$ can bring 
nuclei into thermal equilibrium with the electrons and dramatically
increase the apparent specific heat of the 2D electron 
system\cite{bayot}. We speculate that the sharp peak in the apparent 
specific heat observed by Bayot {\it et al.}\cite{bayot} may be
associated with critical slowing down at the skyrmion melting temperature 
which suppresses motional narrowing of the nuclear resonance in the 
well and thereby enhances thermal coupling to nuclei outside the 
quantum wells.\cite{barunpub} If so, the skyrmion melting
temperature at $|\nu -1 | \sim 0.2$ is $\sim 30-40\ {\rm mK}$,
a believable $\sim10$ times smaller than the naive KTHNY estimate. 

This work was supported in part by NATO Collaborative Research Grant No.
930684, by the National Science Foundation under grants DMR-9416906, and
DMR-9503814 and by CICyT of Spain under contract MAT 94-16906. HAF
acknowledges the support of the A.P. Sloan Foundation and the Research
Corporation. Helpful conversations with S. Barrett, Matthew Fisher,
E. Fradkin, L. Martin-Moreno, K. Moon, 
J. Palacios, N. Read, S. Sachdev, S. Sondhi, 
C. Tejedor, and K. Yang are gratefully acknowledged.

\begin{figure}
\caption{Collective mode energies in $e^2/(\epsilon \ell)$ units
for $\nu=1.10,\tilde g = 0.016 (e^2/\epsilon \ell)$, along the
direction $\Gamma-X$. The shaded region is the magnetic Brillouin
zone.  
At each ${\bf k}$ a 
mode is labeled by the response function component with 
the largest residue in $\chi({\bf k},{\bf k})$.}
\label{collective}
\end{figure}

\begin{figure}
\caption{Parameters of the low energy effective model for 
$\nu = 1.05$ (diamonds), $\nu = 1.1$ (plusses), $\nu = 1.15$ (circles),
and $\nu = 1.2$ (squares)
as a function of $\tilde g/(e^2/\epsilon \ell)$. In the right
panel the solid symbols give values of the boson interaction parameter
$U$ while the open symbols give values of the boson
hopping parameter $J$.
The noise in the latter curve reflects numerical uncertainty introduced by 
the fitting procedure used to extract the spin-wave velocity
from calculated collective mode energies.}
\label{JUKfig}
\end{figure}

\begin{figure}
\caption{$T=0$ phase diagram for Skyrme crystal states.  
The value of $ \tilde g$ at which the square lattice 
shear modulus vanishes has been determined numerically 
at $|\nu-1| = 0.05, 0.1, 0.15,$ and $0.2$ and interpolated to
delimit the square lattice stability region.  
The solid circles indicate the variation of $K_0=\langle K \rangle$ 
along the square lattice region boundary.  Non-collinear 
magnetic order survives quantum fluctuations in the region (approximately)
indicated by the shading.
}
\label{phasediagram}
\end{figure}

\end{document}